\documentclass[10pt,a4paper,onecolumn]{article}
\usepackage{marginnote}
\usepackage{graphicx}
\usepackage{xcolor}
\usepackage{authblk,etoolbox}
\usepackage{titlesec}
\usepackage{calc}
\usepackage{tikz}
\usepackage{hyperref}
\hypersetup{colorlinks,breaklinks,
            urlcolor=[rgb]{0.0, 0.5, 1.0},
            linkcolor=[rgb]{0.0, 0.5, 1.0}}
\usepackage{caption}
\usepackage{tcolorbox}
\usepackage{amssymb,amsmath}
\usepackage{ifxetex,ifluatex}
\usepackage{seqsplit}
\usepackage{fixltx2e} 
\usepackage[
  backend=biber,
]{biblatex}

\usepackage[top=3.5cm, bottom=3cm, right=1.5cm, left=1.0cm,
            headheight=2.2cm, reversemp, includemp, marginparwidth=4.5cm]{geometry}

\titleformat{\section}
  {\normalfont\sffamily\Large\bfseries}
  {}{0pt}{}
\titleformat{\subsection}
  {\normalfont\sffamily\large\bfseries}
  {}{0pt}{}
\titleformat{\subsubsection}
  {\normalfont\sffamily\bfseries}
  {}{0pt}{}
\titleformat*{\paragraph}
  {\sffamily\normalsize}

\usepackage{fancyhdr}
\makeatletter
\let\ps@plain\ps@fancy
\fancyheadoffset[L]{4.5cm}
\fancyfootoffset[L]{4.5cm}

\definecolor{linky}{rgb}{0.0, 0.5, 1.0}

\newtcolorbox{repobox}
   {colback=red, colframe=red!75!black,
     boxrule=0.5pt, arc=2pt, left=6pt, right=6pt, top=3pt, bottom=3pt}

\newcommand{\ExternalLink}{%
   \tikz[x=1.2ex, y=1.2ex, baseline=-0.05ex]{%
       \begin{scope}[x=1ex, y=1ex]
           \clip (-0.1,-0.1)
               --++ (-0, 1.2)
               --++ (0.6, 0)
               --++ (0, -0.6)
               --++ (0.6, 0)
               --++ (0, -1);
           \path[draw,
               line width = 0.5,
               rounded corners=0.5]
               (0,0) rectangle (1,1);
       \end{scope}
       \path[draw, line width = 0.5] (0.5, 0.5)
           -- (1, 1);
       \path[draw, line width = 0.5] (0.6, 1)
           -- (1, 1) -- (1, 0.6);
       }
   }

\patchcmd{\@maketitle}{center}{flushleft}{}{}
\patchcmd{\@maketitle}{center}{flushleft}{}{}
\patchcmd{\@maketitle}{\LARGE}{\LARGE\sffamily}{}{}
\def\maketitle{{%
  
  \AB@maketitle}}
\makeatletter
\renewcommand\AB@affilsepx{ \protect\Affilfont}
\renewcommand\AB@affilnote[1]{{\bfseries #1}\hspace{3pt}}
\makeatother

\renewcommand\Affilfont{\sffamily\small\mdseries}
\ifnum 0\ifxetex 1\fi\ifluatex 1\fi=0 
  \usepackage[T1]{fontenc}
  \usepackage[utf8]{inputenc}

\else 
  \ifxetex
    \usepackage{mathspec}
  \else
    \usepackage{fontspec}
  \fi
  \defaultfontfeatures{Ligatures=TeX,Scale=MatchLowercase}

\fi
\IfFileExists{upquote.sty}{\usepackage{upquote}}{}
\IfFileExists{microtype.sty}{%
\usepackage{microtype}
\UseMicrotypeSet[protrusion]{basicmath} 
}{}

\usepackage{hyperref}
\hypersetup{unicode=true,
            pdftitle={proEQUIB: IDL Library for Plasma Diagnostics and Abundance Analysis},
            pdfborder={0 0 0},
            breaklinks=true}
\usepackage{graphicx,grffile}
\makeatletter
\def\maxwidth{\ifdim\Gin@nat@width>\linewidth\linewidth\else\Gin@nat@width\fi}
\def\maxheight{\ifdim\Gin@nat@height>\textheight\textheight\else\Gin@nat@height\fi}
\makeatother
\setkeys{Gin}{width=\maxwidth,height=\maxheight,keepaspectratio}
\IfFileExists{parskip.sty}{%
\usepackage{parskip}
}{
\setlength{\parindent}{0pt}
\setlength{\parskip}{6pt plus 2pt minus 1pt}
}
\ifx\paragraph\undefined\else
\let\oldparagraph\paragraph
\renewcommand{\paragraph}[1]{\oldparagraph{#1}\mbox{}}
\fi
\ifx\subparagraph\undefined\else
\let\oldsubparagraph\subparagraph
\renewcommand{\subparagraph}[1]{\oldsubparagraph{#1}\mbox{}}
\fi

\title{proEQUIB: IDL Library for Plasma Diagnostics and Abundance Analysis}

        \author[1, 2]{Ashkbiz Danehkar}
    
      \affil[1]{Research Centre in Astronomy, Astrophysics and Astrophotonics, Macquarie University, Sydney, NSW 2109, Australia}
      \affil[2]{Harvard-Smithsonian Center for Astrophysics, 60 Garden Street, Cambridge, MA 02138, USA}
  \date{\vspace{-5ex}}

\begin{document}
\maketitle

\marginpar{
  \sffamily\small

  {\bfseries DOI:} \href{https://doi.org/10.21105/joss.00899}{\color{linky}{10.21105/joss.00899}}

  \vspace{2mm}

  {\bfseries Software}
  \begin{itemize}
    \setlength\itemsep{0em}
    \item \href{https://github.com/openjournals/joss-reviews/issues/899}{\color{linky}{Review}} \ExternalLink
    \item \href{https://github.com/equib/proEQUIB}{\color{linky}{Repository}} \ExternalLink
    \item \href{https://doi.org/10.5281/zenodo.1890337}{\color{linky}{Archive}} \ExternalLink
  \end{itemize}

  \vspace{2mm}

  {\bfseries Submitted:} 30 June 2018\\
  {\bfseries Published:} 04 December 2018

  \vspace{2mm}
  {\bfseries License}\\
  Authors of papers retain copyright and release the work under a Creative Commons Attribution 4.0 International License (\href{http://creativecommons.org/licenses/by/4.0/}{\color{linky}{CC-BY}}).
}

\vspace{8mm}
  
\hypertarget{summary}{%
\section{Summary}\label{summary}}

The emission lines emitted from gaseous nebulae carry valuable
information about the physical conditions and chemical abundances of
ionized gases in these objects, as well as the interstellar reddening.
We determine the electron temperature, the electron density, and the
ionic abundances from the dereddened fluxes of \emph{collisionally
excited lines} (CEL) and \emph{recombination lines} (RL) identified in
nebular spectra (see e.g. Danehkar, Parker, and Ercolano (2013);
Danehkar et al. (2014); Danehkar, Parker, and Steffen (2016); Danehkar
(2018b)).

\texttt{proEQUIB} is a library including several application programming
interface (API) functions developed in the Interactive Data Language
(IDL), which can be used to determine temperatures, densities, and
chemical abundances from emission lines of ionized nebulae. This IDL
library can also be used by the GNU Data Language (GDL) (Arabas et al.
(2010); Coulais et al. (2010)), which is a free and open-source
alternative IDL compiler. This IDL/GDL package employs the IDL library
\texttt{AtomNeb} \emph{Atomic Data for Ionized Nebulae} (Danehkar
(2018a)), which contains collision strengths and transition
probabilities for collisional excitation calculations, and recombination
coefficients for recombination calculations. This package includes
several API functions to determine physical conditions and chemical
abundances from CEL and RL, derive interstellar extinctions from Balmer
lines, and deredden the observed fluxes:

\begin{itemize}
\item
  The API functions for the \emph{CEL analysis} were developed in the
  IDL programming language based on the algorithm of the FORTRAN program
  \texttt{EQUIB} (Howarth and Adams (1981); Howarth et al. (2016)),
  which calculates atomic level populations and line emissivities in
  statistical equilibrium in multi-level atoms for the given physical
  conditions. These API functions can be used to determine the electron
  temperature, the electron density, and the ionic abundances from the
  dereddened fluxes of \emph{collisionally excited lines} emitted from
  ionized gaseous nebulae.
\item
  The API functions for the \emph{RL analysis} were developed in IDL
  according to the algorithm of the recombination scripts by X. W. Liu
  and Y. Zhang included in the FORTRAN program \texttt{MOCASSIN}
  (Ercolano et al. (2003); Ercolano, Barlow, and Storey (2005)). These
  API functions can be used to determine the ionic abundances from the
  dereddened fluxes of \emph{recombination lines} emitted from ionized
  nebulae.
\item
  The API functions for the \emph{reddening analysis} were developed
  based on the methods of the reddening functions from the Space
  Telescope Science Data Analysis System (\texttt{STSDAS}) IRAF Package
  (Bushouse and Simon (1994); Shaw and Dufour (1994)). These API
  functions can be employed to obtain \emph{interstellar extinctions}
  for different reddening laws from the observed fluxes of Balmer lines
  detected in nebular spectra, and deredden the measured fluxes of
  emission lines.
\end{itemize}

\texttt{proEQUIB} has recently been used for plasma diagnostics and
abundance analysis of some planetary nebulae (Danehkar, Parker, and
Steffen (2016); Danehkar (2018b)). This IDL/GDL package heavily relies
on the IDL Astronomy User's library (Landsman (1993); Landsman (1995))
and the IDL library \texttt{AtomNeb} (Danehkar (2018a)). The API
functions of this IDL library can easily be utilized to generate
spatially-resolved maps of extinction, temperature, density, and
chemical abundances from integral field spectroscopic observations (see
e.g. Danehkar, Parker, and Ercolano (2013); Danehkar et al. (2014);
Danehkar (2014)).

\hypertarget{acknowledgements}{%
\section{Acknowledgements}\label{acknowledgements}}

A.D. acknowledges the receipt of a Macquarie University Research
Excellence Scholarship.

\hypertarget{references}{%
\section*{References}\label{references}}
\addcontentsline{toc}{section}{References}

\hypertarget{refs}{}
\leavevmode\hypertarget{ref-Arabas:2010}{}%
Arabas, S., M. Schellens, A. Coulais, J. Gales, and P. Messmer. 2010.
``GNU Data Language (GDL) - a free and open-source implementation of
IDL.'' In \emph{EGU General Assembly Conference}, 12:924. Geophysical
Research Abstracts.

\leavevmode\hypertarget{ref-Bushouse:1994}{}%
Bushouse, H., and B. Simon. 1994. ``The IRAF/STSDAS Synthetic Photometry
Package.'' In \emph{Astronomical Data Analysis Software and Systems
Iii}, edited by D. R. Crabtree, R. J. Hanisch, and J. Barnes, 61:339.
Astronomical Society of the Pacific Conference Series.

\leavevmode\hypertarget{ref-Coulais:2010}{}%
Coulais, A., M. Schellens, J. Gales, S. Arabas, M. Boquien, P. Chanial,
P. Messmer, et al. 2010. ``Status of GDL - GNU Data Language.'' In
\emph{Astronomical Data Analysis Software and Systems Xix}, edited by Y.
Mizumoto, K.-I. Morita, and M. Ohishi, 434:187. Astronomical Society of
the Pacific Conference Series.

\leavevmode\hypertarget{ref-Danehkar:2014b}{}%
Danehkar, A. 2014. ``Evolution of Planetary Nebulae with WR-type Central
Stars.'' PhD thesis, Macquarie University, Australia.
doi:\href{https://doi.org/10.5281/zenodo.47794}{\color{linky}{10.5281/zenodo.47794}}.

\leavevmode\hypertarget{ref-Danehkar:2018b}{}%
Danehkar, A. 2018a. ``AtomNeb: Atomic Data for Ionized Nebulae.''
\emph{The Journal of Open Source Software} submitted.

\leavevmode\hypertarget{ref-Danehkar:2018}{}%
Danehkar, A. 2018b. ``Bi-Abundance Ionisation Structure of the Wolf-Rayet
Planetary Nebula PB 8.'' \emph{Publications of the Astronomical Society
of Australia} 35: e005. 
doi:\href{https://doi.org/10.1017/pasa.2018.1}{\color{linky}{10.1017/pasa.2018.1}}.

\leavevmode\hypertarget{ref-Danehkar:2013}{}%
Danehkar, A., Q. A. Parker, and B. Ercolano. 2013. ``Observations and
three-dimensional ionization structure of the planetary nebula SuWt 2.''
\emph{Monthly Notices of the Royal Astronomical Society} 434: 1513--30.
doi:\href{https://doi.org/10.1093/mnras/stt1116}{\color{linky}{10.1093/mnras/stt1116}}.

\leavevmode\hypertarget{ref-Danehkar:2016}{}%
Danehkar, A., Q. A. Parker, and W. Steffen. 2016. ``Fast, Low-ionization
Emission Regions of the Planetary Nebula M2-42.'' \emph{The Astronomical
Journal} 151: 38. 
doi:\href{https://doi.org/10.3847/0004-6256/151/2/38}{\color{linky}{10.3847/0004-6256/151/2/38}}.

\leavevmode\hypertarget{ref-Danehkar:2014}{}%
Danehkar, A., H. Todt, B. Ercolano, and A. Y. Kniazev. 2014.
``Observations and three-dimensional photoionization modelling of the
Wolf-Rayet planetary nebula Abell 48.'' \emph{Monthly Notices of the
Royal Astronomical Society} 439: 3605--15. \\
doi:\href{https://doi.org/10.1093/mnras/stu203}{\color{linky}{10.1093/mnras/stu203}}.

\leavevmode\hypertarget{ref-Ercolano:2005}{}%
Ercolano, B., M. J. Barlow, and P. J. Storey. 2005. ``The dusty
MOCASSIN: fully self-consistent 3D photoionization and dust radiative
transfer models.'' \emph{Monthly Notices of the Royal Astronomical
Society} 362: 1038--46.
doi:\href{https://doi.org/10.1111/j.1365-2966.2005.09381.x}{\color{linky}{10.1111/j.1365-2966.2005.09381.x}}.

\leavevmode\hypertarget{ref-Ercolano:2003}{}%
Ercolano, B., M. J. Barlow, P. J. Storey, and X.-W. Liu. 2003.
``MOCASSIN: a fully three-dimensional Monte Carlo photoionization
code.'' \emph{Monthly Notices of the Royal Astronomical Society} 340:
1136--52. 
doi:\href{https://doi.org/10.1046/j.1365-8711.2003.06371.x}{\color{linky}{10.1046/j.1365-8711.2003.06371.x}}.

\leavevmode\hypertarget{ref-Howarth:1981}{}%
Howarth, I. D., and S. Adams. 1981. ``Program EQUIB.'' University
College London.

\leavevmode\hypertarget{ref-Howarth:2016}{}%
Howarth, I. D., S. Adams, R. E. S. Clegg, D. P. Ruffle, X.-W. Liu, C. J.
Pritchet, and B. Ercolano. 2016. ``EQUIB: Atomic level populations and
line emissivities calculator.'' Astrophysics Source Code Library,
\href{http://ascl.net/1603.005}{\color{linky}{ascl:1603.005}}.

\leavevmode\hypertarget{ref-Landsman:1993}{}%
Landsman, W. B. 1993. ``The IDL Astronomy User's Library.'' In
\emph{Astronomical Data Analysis Software and Systems Ii}, edited by R.
J. Hanisch, R. J. V. Brissenden, and J. Barnes, 52:246. Astronomical
Society of the Pacific Conference Series.

\leavevmode\hypertarget{ref-Landsman:1995}{}%
Landsman, W. B. 1995. ``The IDL Astronomy User's Library.'' In
\emph{Astronomical Data Analysis Software and Systems Iv}, edited by R.
A. Shaw, H. E. Payne, and J. J. E. Hayes, 77:437. Astronomical Society
of the Pacific Conference Series.

\leavevmode\hypertarget{ref-Shaw:1994}{}%
Shaw, R. A., and R. J. Dufour. 1994. ``The FIVEL Nebular Modelling
Package in STSDAS.'' In \emph{Astronomical Data Analysis Software and
Systems Iii}, edited by D. R. Crabtree, R. J. Hanisch, and J. Barnes,
61:327. Astronomical Society of the Pacific Conference Series.

\end{document}